\RequirePackage{lineno}
\documentclass[twocolumn,superscriptaddress,preprintnumbers,amsmath,amssymb,prc]{revtex4}  
\usepackage{graphicx}
\usepackage{dcolumn}
\usepackage{bm}
\usepackage{float}
\usepackage{color}
\usepackage{CJK}
\usepackage[colorlinks,linkcolor=blue,urlcolor=blue,citecolor=blue]{hyperref}
\usepackage{isotope}
\usepackage[normalem]{ulem}
\renewcommand{\sout}{\bgroup \color{red} \ULdepth=-0.5ex \ULset}

\usepackage{amsmath}
\usepackage{cases}
\usepackage{amssymb}
\usepackage{eqnalign}

\begin{document}
\begin{CJK*}{UTF8}{gbsn}

\title{Difference between signal and background of the chiral magnetic effect relative to spectator and participant planes in isobar collisions at $\sqrt{s_{_{\rm NN}}} = 200$ GeV}

\author{Bang-Xiang Chen}
\affiliation{Key Laboratory of Nuclear Physics and Ion-beam Application~(MOE), Institute of Modern Physics, Fudan University, Shanghai $200433$, China}
\affiliation{Shanghai Research Center for Theoretical Nuclear Physics, NSFC and Fudan University, Shanghai $200438$, China}

\author{Xin-Li Zhao}
\affiliation{College of Science, University of Shanghai for Science and Technology, Shanghai $200093$, China}
\affiliation{Key Laboratory of Nuclear Physics and Ion-beam Application~(MOE), Institute of Modern Physics, Fudan University, Shanghai $200433$, China}
\affiliation{Shanghai Research Center for Theoretical Nuclear Physics, NSFC and Fudan University, Shanghai $200438$, China}

\author{Guo-Liang Ma}
\email{glma@fudan.edu.cn}
\affiliation{Key Laboratory of Nuclear Physics and Ion-beam Application~(MOE), Institute of Modern Physics, Fudan University, Shanghai $200433$, China}
\affiliation{Shanghai Research Center for Theoretical Nuclear Physics, NSFC and Fudan University, Shanghai $200438$, China}


\begin{abstract}
The search for the chiral magnetic effect (CME) in relativistic heavy-ion collisions helps us understand the $\mathcal{CP}$ symmetry breaking in strong interactions and the topological nature of the QCD vacuum. Since the background and signal of the CME have different correlations with the spectator and participant planes, a two-plane method has been proposed to extract the fraction of the CME signal inside the CME observable of $\Delta\gamma$ from the experimental measurements relative to the two planes. Using a multiphase transport model with different strengths of the CME, we reexamine the two-plane method in isobar collisions at $\sqrt{s_{_{\rm NN}}} = 200$ GeV. The ratios of the CME signals and the elliptic flow backgrounds relative to the two different planes are found to be different, which is inconsistent with the assumptions made in the current experimental measurements. This difference arises from the decorrelation effect of the chiral magnetic effect relative to the spectator and participant planes caused by final state interactions. Our finding suggests that the current experimental measurements may overestimate the fraction of the CME signal in the CME observable in the final state of relativistic heavy-ion collisions.
\end{abstract}


\maketitle

\section{Introduction}
\label{sec:intro}
Relativistic heavy-ion collisions not only produce the quark-gluon plasma with strong collectivity~\cite{Kolb:2000sd,Teaney:2000cw,Yan:2017ivm,Shen:2020mgh,Song:2017wtw,Lan:2022rrc,Wu:2021xgu,Wang:2022fwq,Wang:2022det}, but also generate the strongest magnetic field as the spectator protons from the target and the projectile pass through each other at almost the speed of light~\cite{Skokov:2009qp,Bzdak:2011yy,Deng:2012pc,Zhao:2017rpf,Zhao:2019ybo,Chen:2021nxs}. This provides a unique experimental way to study the topological properties of the QCD vacuum and the anomalous chiral transport phenomena under the strong magnetic field~\cite{Kharzeev:2020jxw,Huang:2015oca,Hattori:2016emy,Gao:2020vbh,Xin-Li:2023hqa}. One of the possible experimental probes is through the chiral magnetic effect (CME), which would lead to charge separation along the direction of the magnetic field in a system with chiral imbalance~\cite{Kharzeev:2004ey,Kharzeev:2007jp,Fukushima:2008xe}.

The charge-dependent azimuthal correlation was first proposed as a possible observable for detecting the CME~\cite{Voloshin:2004vk}, i.e., $\gamma_{\alpha\beta}=\langle \cos(\phi_{\alpha}+\phi_{\beta}-2\Psi_{\rm RP})\rangle$, where $\phi_{\alpha(\beta)}$ is the azimuthal angle of a charged particle $\alpha(\beta)$, and $\Psi_{\rm RP}$ is the angle of the reaction plane, and $\Delta\gamma$ denotes the difference between opposite-charged and same-charged correlations. The early measurements of the charge-dependent azimuthal correlations by the STAR Collaboration~\cite{STAR:2009wot,STAR:2009tro,STAR:2013ksd,STAR:2014uiw}  at the BNL Relativistic Heavy Ion Collider (RHIC) and the ALICE Collaboration~\cite{ALICE:2012nhw} at the CERN Large Hadron Collider (LHC) were consistent with the CME expectations. Unfortunately, the background effect significantly affects the measured correlations due to the presence of strong collective flow, especially from elliptical flow~\cite{Bzdak:2010fd,Liao:2010nv,Schlichting:2010qia,Wang:2009kd,Wu:2022fwz,STAR:2013zgu}. The recent RHIC-STAR measurements gave a strict constraint that the CME fraction extracted in Au+Au 200 GeV is very small, less than 10\%~\cite{Wang:2018ygc,Zhao:2020utk,Zhao:2019hta,Wang:2022eoo}. In order to distinguish the possible CME signal from the dominant background, many different methods or schemes have been proposed~\cite{Zhao:2019hta,Wang:2022eoo,Li:2020dwr,Qi-Ye:2023zyf}. One of the most important schemes is to use isobar collisions because the two isobar systems ($_{44}^{96}\textrm{Ru}+_{44}^{96}\textrm{Ru}$ and $_{40}^{96}\textrm{Zr}+_{40}^{96}\textrm{Zr}$) have the same nucleon number, but different proton numbers~\cite{Voloshin:2010ut}.  Theoretically, it was expected that in a similar elliptical flow-induced background, there might be a $20\%$ difference in their CME observables~\cite{Deng:2016knn,Koch:2016pzl,Deng:2018dut,Choudhury:2021jwd}. This led the RHIC-STAR Collaboration to perform the isobar collision experiments on $_{44}^{96}\textrm{Ru}+_{44}^{96}\textrm{Ru}$ and $_{40}^{96}\textrm{Zr}+_{40}^{96}\textrm{Zr}$ collisions at $\sqrt{s_{_{\rm NN}}} = 200$ GeV~\cite{STAR:2019bjg,STAR:2021mii}.

Since the CME signal is positively correlated with the magnetic field strength, the ratio of the CME observable from Ru+Ru collisions to that from Zr+Zr collisions is theoretically predicted to be greater than 1. However, the recent experimental results published by STAR observed that the ratios of the various CME observables are less than unity~\cite{STAR:2021mii}. This suggests that the background effect is more dominant than the CME signal, and that the CME signal is either absent or very small in isobaric collisions. How to understand the results of the isobar collision experiment has recently become a research direction of great interest. Different nuclear deformations or nuclear structures~\cite{Butler:1996zz,Zhou:2023vgv,Rong:2022qez,Shi:2021far,Shang:2022ntl} have been used to explain the differences in multiplicity and harmonic flows between the two isobar systems~\cite{Xu:2021vpn,Xu:2021uar,Zhang:2021kxj,Jia:2021oyt,Jia:2021tzt,Jia:2021qyu,Xu:2017zcn,Li:2019kkh}. Taking into account both the halo-type neutron skin structure and CME-like charge separation, we have shown that it is difficult to use the CME observables to distinguish the presence or absence of the CME if the CME strength is weak in isobar collisions~\cite{Zhao:2022grq,Feng:2021oub}. Meanwhile, based on the recent finding from the anomalous viscous fluid dynamics (AVFD) model, the STAR results favor a limited CME signal contribution of about $(6.8 \pm 2.6)\%$~\cite{Kharzeev:2022hqz}.The latest STAR evaluation of the CME signal extracts an upper limit for the CME fraction in the $\Delta\gamma$ measurement of approximately $10\%$ at a $95\%$ confidence level in isobar collisions at $\sqrt{s_{_{\rm NN}}} = 200$ GeV after considering nonflow contamination~\cite{STAR:2023gzg,STAR:2023ioo}.
Many experimental observables have been used to probe the CME in Au+Au and isobar collisions. A two-plane measurement method that utilizes the charge-dependent azimuthal correlations relative to the spectator plane (SP) and participant plane (PP) has been proposed in Refs.~\cite{Xu:2017qfs,Xu:2017zcn}, because the background and the CME signal have different sensitivities or correlations to the two planes~\cite{Zhao:2019crj}. The STAR Collaboration has used the method to detect the fraction of the CME signal inside the inclusive $\Delta\gamma$ correlation in both Au+Au and isobar collisions. For Au + Au collision at $\sqrt{s_{_{\rm NN}}} = 200$ GeV, the STAR results indicate that the fraction of the CME signal is consistent with zero in peripheral centrality bins, but that there may be a finite CME signal in mid-central centrality bins~\cite{STAR:2021pwb}. This method is believed to eliminate most of the contribution from the collective flow in the background effect, but further study is needed regarding how to deduct some of the nonflow background effects~\cite{Feng:2021pgf}.

Regarding the spectator and participant plane methodology, it is assumed that the ratio $a$ of elliptic flow relative to different reaction planes is as same as the ratio $b$ of CME signals relative to different reaction planes~\cite{STAR:2021pwb,Shi:2017cpu,Feng:2021pgf}. However, it is possible that the two ratios are different. It is thus essential to theoretically study the ratio between the CME signals relative to different reaction planes. This motivates us to calculate the ratios of $a$ and $b$ in isobar collisions at $\sqrt{s_{_{\rm NN}}} = 200$ GeV using a multiphase transport (AMPT) model with an initial CME signal, in order to provide some theoretical support for experimental measurements of the CME.

The paper is organized as follows. In Sec.~\ref{sec:model}, we introduce the framework of the AMPT model with an initial CME signal for isobar collisions and how we extract the fraction of the CME signal contained in the inclusive $\Delta\gamma$ using the two-plane method. In Sec.~\ref{sec:results}, we compare our model results with the measurements from the STAR experiment and discuss the implications of our findings for interpreting experimental data, as well as the physical sources from which our results arise. Finally, a summary is provided in Sec.~\ref{sec:summary}.

\section{Model and method}
\label{sec:model}
\subsection{The AMPT model with initial CME signal}

The AMPT model is a hybrid transport model that contains the following four packages for modeling the four main stages of relativistic heavy-ion collisions~\cite{Lin:2004en,Ma:2016fve,Lin:2021mdn}. (1) The HIJING model provides the following initial conditions. The transverse density distribution of colliding nuclei is considered as a Woods-Saxon distribution. The multiple scatterings between participating nuclei produce the spatial and momentum space distributions of minijet partons and soft excited strings. With the help of a string melting mechanism, the quark plasma is produced by melting the parent hadrons. (2) Zhang's parton cascade model is used to simulate the stage of parton cascade. The ZPC model describes Parton interactions for two-body elastic scatterings, where the parton cross section is computed from the leading-order perturbative QCD (pQCD) calculation for gluon-gluon elastic scatterings. (3) A quark coalescence model combines two or three nearest partons into hadrons in order to simulate the hadronization.  (4)  A relativistic transport (ART) model simulates the stage of hadronic rescatterings, including all hadronic reaction channels for elastic and inelastic scatterings of baryon-baryon, baryon-meson, and meson-meson interactions, as well as resonance decays. Many previous studies have shown that the AMPT model provides a good description of a wide range of experimental observables in both large and small colliding systems at RHIC and the LHC~\cite{Lin:2004en,Ma:2016fve,Lin:2021mdn,Lin:2014tya,OrjuelaKoop:2015jss,Ma:2016bbw,He:2017tla,Huang:2021ihy,Chen:2022wkj,Chen:2022xpm}.

To simulate isobar collisions, the spatial distributions of nucleons inside $_{44}^{96}\textrm{Ru}$ and $_{40}^{96}\textrm{Zr}$ in the rest frame are sampled according to the following Woods-Saxon form in the spherical coordinate system,
\begin{eqnarray}
\rho (r,\theta )&=&\rho _{0}/\{1+\exp[(r-R(\theta,\phi))/a_{0}]\}, \\
R(\theta ,\phi )&=&R_{0}[1+\beta_{2} Y_{2,0}(\theta,\phi)+\beta_{3} Y_{3,0}(\theta,\phi)],
\label{rho}
\end{eqnarray}%
where $\rho _{0}$ is the normal nuclear density, $a_{0}$ is the surface diffuseness parameter, $R_{0}$ is the nucleus radius, and $\beta _{2}$ and $\beta _{3}$ are the quadrupole and octupole deformities for the nucleus. In our previous work~\cite{Zhao:2022grq}, we found that the halo-type neutron skin case~\cite{Xu:2021vpn} is the best of the eighteen cases because it can simultaneously the experimental ratio of charged-particle multiplicity distributions, the average number of charged particles, and elliptic flow. Therefore, we also choose the halo-type neutron skin case in this study. In this case, i.e., no deformation for both $_{44}^{96}\textrm{Ru}$ and $_{40}^{96}\textrm{Zr}$ ($\beta_2=\beta_3=0$), $R_0=5.085$ and $a_{0}=0.523$ for both protons and neutrons inside $_{44}^{96}\textrm{Ru}$, but $R_0=5.021$ and $a_{0}=0.523$ for protons and $R_0=5.021$ and $a_{0}=0.592$ for neutrons inside $_{40}^{96}\textrm{Zr}$ because of the possible presence of a neutron halo in $_{40}^{96}\textrm{Zr}$.

We introduce a CME-like charge separation in the initial partonic stage of the AMPT model using the approach in Ref.~\cite{Ma:2011uma}. By adjusting the percentage $p$ that defines the percentage of quarks joining the charge separation, we can control the signal strength of the CME. The $p$ is defined as follows,
\begin{equation}
p = \frac{N_{\uparrow(\downarrow)}^{+(-)}-N_{\downarrow(\uparrow)}^{+(-)}}{N_{\uparrow(\downarrow)}^{+(-)}+N_{\downarrow(\uparrow)}^{+(-)}},
 \label{eq-f}
\end{equation}
where $N$ is the number of quarks of a given species (u or d or s), $+$ and $-$ denote positive and negative charges of quarks, and $\uparrow$ and $\downarrow$ denote the directions in which the quarks move along the magnetic field.
Considering that the magnetic fields of $\rm Ru+Ru$ and $\rm Zr+Zr$ collisions differ at the event level ~\cite{Zhao:2019crj}, we actually perform the initial charge separation based on the magnitude and direction of the magnetic field for each event. To simplify the expression, we denote $p$ as the CME strength in Ru+Ru collisions, e.g., $p=2\%$ means that $p_{\rm Ru+Ru}=2\%$ and $p_{\rm Zr+Zr}=2\%/1.15=1.74\%$ since we keep $p_{\rm Ru+Ru}/p_{\rm Zr+Zr}=1.15$.

\subsection{Spectator and participant planes}
\label{sec:plane}
In the two-plane method, the elliptic flow-driven background is believed to be more correlated with the participant plane (PP), but the CME signal is more correlated with the spectator plane (SP)~\cite{Xu:2017qfs,Xu:2017zcn}. We reconstruct the spectator and participant planes, respectively, using the two equations,

\begin{equation}\label{equ.01}
	\psi_{\rm S P}=\frac{\operatorname{atan} 2\left(\left\langle r_{\rm n}^2 \sin \left(2 \phi_{\text {n }}\right)\right\rangle,\left\langle r_{\rm n}^2 \cos \left(2 \phi_{\text {n }}\right)\right\rangle\right)}{2},
\end{equation}
\begin{equation}\label{equ.02}
	\psi_{\rm P P}=\frac{\operatorname{atan} 2\left(\left\langle r_{\rm p}^2 \sin \left(2 \phi_{\text {p }}\right)\right\rangle,\left\langle r_{\rm p}^2 \cos \left(2 \phi_{\text {p }}\right)\right\rangle\right)+\pi}{2},
\end{equation}
where $r_{\rm n}$ and $\phi_{\rm n}$ are the displacement and azimuthal angle of spectator neutrons in the transverse plane, and $r_{\rm p}$ and $\phi_{\rm p}$ are the displacement and azimuthal angle of participating partons in the transverse plane, respectively. All spatial information on the displacements and azimuthal angles is obtained from the initial state of the AMPT model. Note that the PP can be experimentally assessed by the event plane reconstructed from final-state hadrons. To avoid the nonflow effect in the reconstructed $\Psi_{\mathrm{EP}}$~\cite{Feng:2021pgf}, we have chosen PP for our calculations in this study. However, we have verified that our conclusions do not change in the EP case. With the two different planes,  the corresponding elliptic flows $v_2\{\mathrm{SP}\}$ and $v_2\{\mathrm{PP}\}$ can be calculated, respectively, as
\begin{equation}\label{equ.03}
	v_2\{\rm SP\}=\left\langle\cos 2\left(\phi-\psi_{\mathrm{SP}}\right)\right\rangle,
\end{equation}
\begin{equation}\label{equ.04}
	v_2\{\rm PP\}=\left\langle\cos 2\left(\phi-\psi_{\mathrm{PP}}\right)\right\rangle.
\end{equation}

\begin{figure}
	\includegraphics[scale=0.4]{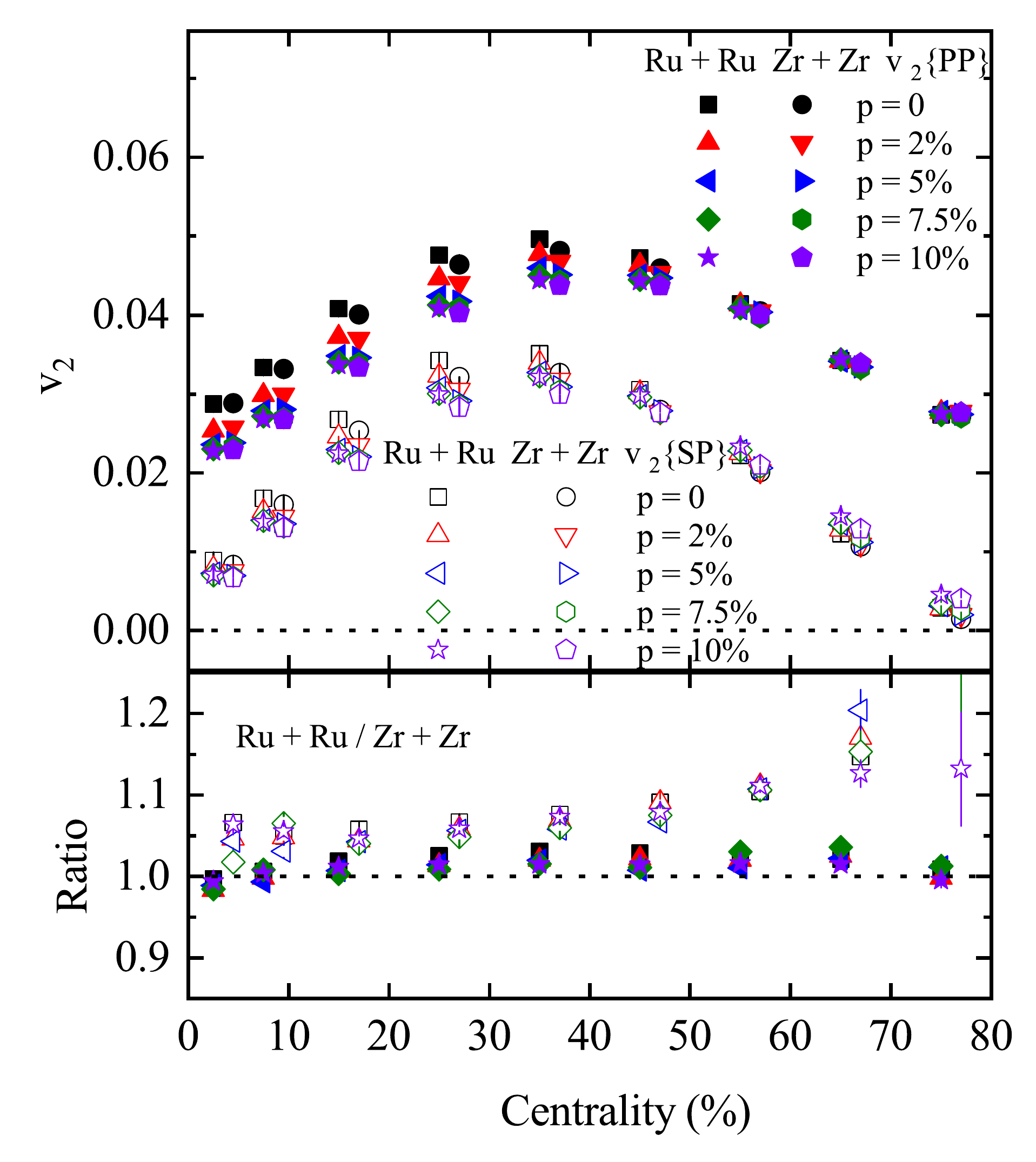}
	\caption{(Color online) Upper panel: AMPT results on centrality dependences of elliptic flow $v_2\{\mathrm{PP}\}$ (solid symbols) and $v_2\{\mathrm{SP}\}$ (open symbols) in isobar collisions at $\sqrt{s_{_{\rm NN}}} = 200$ GeV from the AMPT model with different strengths of the CME. Lower panel: AMPT results on $v_2\{\mathrm{PP}\}$ ratios and $v_2\{\mathrm{SP}\}$ ratios of Ru + Ru collisions to Zr + Zr collisions as a function of centrality bin.}
	\label{fig-v2}
\end{figure}

The upper panel in Fig.~\ref{fig-v2} shows the centrality dependences of $v_2\{\mathrm{PP}\}$ and $v_2\{\mathrm{SP}\}$ of charged hadrons with $0.2<p_T<2.0 ~\mathrm{GeV} / c$ and $|\eta|<1$ from the AMPT model with different strengths of the CME in Ru + Ru and Zr + Zr collisions at $\sqrt{s_{_{\rm NN}}} = 200$ GeV. We can see that $v_2\{\mathrm{PP}\}$ is greater than $v_2\{\mathrm{SP}\}$ in all cases, since elliptic flow is more correlated with the participant plane than with the spectator plane. On the other hand, for the central and mid-central centrality of 0$-$50$\%$, both $v_2\{\mathrm{PP}\}$ and $v_2\{\mathrm{SP}\}$ decrease slightly as  the CME signal strength increases. The $v_2\{\mathrm{PP}\}$ and $v_2\{\mathrm{SP}\}$ are insensitive to the strength of the CME signal for the peripheral centrality bin of 50$-$80$\%$. The lower panel in Fig.~\ref{fig-v2} shows the $v_2\{\mathrm{PP}\}$ and $v_2\{\mathrm{SP}\}$ ratios of Ru + Ru to Zr + Zr. The ratio of $v_2\{\mathrm{SP}\}$ is greater than that of $v_2\{\mathrm{PP}\}$. The trend in the ratio results is caused by the nuclear structures of Ru and Zr~\cite{Zhao:2022grq}.

\subsection{Two-plane method to extract $f_{cme}$}
\label{sec:fcme}
This subsection will first introduce the two-plane method for detecting CME signals and then discuss how the two-plane method can be improved using the AMPT model. The experimentally measured CME observable $\Delta\gamma$ includes the CME signal and the background effect mainly arising from the contributions of elliptical flow and non-flow effects. Therefore, the experimentally measured CME observations $\Delta\gamma$ relative to a plane $\psi$ can be divided into two parts, as follows.
\begin{equation}\label{equ.05}
	\Delta \gamma \{\mathrm{\psi}\}=\Delta \gamma_{\rm Bkg}\{\mathrm{\psi}\}+\Delta \gamma_{\rm CME}\{\mathrm{\psi}\}
\end{equation}
where $\psi$ can be $\psi_{\rm PP}$ or $\psi_{\rm SP}$. The ratios of the elliptic flows and the measured observables relative to two different planes can be defined by $a$ and $A$, respectively, as

\begin{equation}\label{equ.06}
	a=v_{2}\{\mathrm{SP}\} / v_{2}\{\mathrm{PP}\},
\end{equation}
\begin{equation}\label{equ.07}
	A=\Delta \gamma{\{\mathrm{SP}\}} / \Delta \gamma{\{\mathrm{PP}\}}.
\end{equation}

It is expected that $a$ follows a two-plane correlation factor, i.e., $a = \left\langle\cos 2\left(\Psi_{\mathrm{PP}}-\Psi_{\mathrm{SP}}\right)\right\rangle$~\cite{Xu:2017qfs,Bloczynski:2012en,Shi:2019wzi}. Since the CME signal can not be measured directly in experiments, one usually assumes that the ratio of the CME signals relative to different reaction planes is the inverse of $a$. Thus the following relation can be obtained:

\begin{equation}\label{equ.08}
	\Delta \gamma \{\mathrm{SP}\}=a \Delta \gamma_{\rm Bkg}\{\mathrm{PP}\}+\Delta \gamma_{\rm CME}\{\mathrm{PP}\} / a .
\end{equation}

Upon simple solving, the percentage of the CME signal within the measured CME observable (denoted as $f_{\rm CME}$) is obtained by the equation,
\begin{equation}\label{equ.09}
	f_{\rm CME}=\frac{\Delta \gamma_{\rm CME}\{\mathrm{PP}\}}{\Delta \gamma\{\mathrm{PP}\}}=\frac{A / a-1}{1 / a^2-1}.
\end{equation}
Equation~(\ref{equ.09}) shows that the percentage of the CME signal within the measured CME observable can be obtained by measuring $A$ and $a$. However, the assumption made above may not be valid due to a possible difference between the ratio of the CME signals and the inverse ratio of elliptical flows~\cite{Shi:2017cpu,Feng:2021pgf}. Thus, in the more general case the following relationship holds:

\begin{equation}\label{equ.10}
	\Delta \gamma_{\mathrm{CME}}\{\mathrm{PP}\}=b \Delta \gamma_{\mathrm{CME}}\{\mathrm{SP}\}.
\end{equation}
where $b$ represents the ratio of the CME signals relative to the two different planes. A more generalized equation can be obtained by replacing the corresponding part of Eq.~(\ref{equ.08}), as
\begin{equation}\label{equ.11}
	\Delta \gamma \{\mathrm{SP}\}=a \Delta \gamma_{\rm Bkg}\{\mathrm{PP}\}+\Delta \gamma_{\rm CME}\{\mathrm{PP}\} / b .
\end{equation}
Therefore, a more realistic percentage after taking $b$ into account ($f_{\rm CME}\{b\}$) can be calculated as
\begin{equation}\label{equ.12}
	f_{\rm CME}\{b\}=\frac{\Delta \gamma_{\rm CME}\{\mathrm{PP}\}}{\Delta \gamma\{\mathrm{PP}\}}=\frac{A / a-1}{1 / a b-1}.
\end{equation}

The remaining important question is how to calculate $b$, which can be done theoretically. For example, within the theoretical framework of our AMPT model, the value of $b$ can be obtained by the equation
\begin{equation}\label{equ.13}
b =\frac{\Delta \gamma\{\mathrm{PP}\}(p \neq 0) - \Delta \gamma\{\mathrm{PP}\}(p=0)}{\Delta \gamma\{\mathrm{SP}\}(p \neq 0) - \Delta \gamma\{\mathrm{SP}\}(p=0)},
\end{equation}
where the numerator and denominator are the CME signals inside the measured CME observables relative to participant and spectator planes, respectively. They can be obtained by taking the differences in the results with and without the imported CME signal.

\section{Results and Discussions}
\label{sec:results}
In this section, we present the AMPT results on charge-dependent azimuthal correlations for charged particles relative to spectator and participant planes, and compare them with the results from the STAR isobar experiment. We keep the same kinetic cuts as the STAR experiment, i.e., $0.2<p_T<2.0 \mathrm{~GeV} / c$ and $|\eta|<1$.

\begin{figure}
	\includegraphics[scale=0.4]{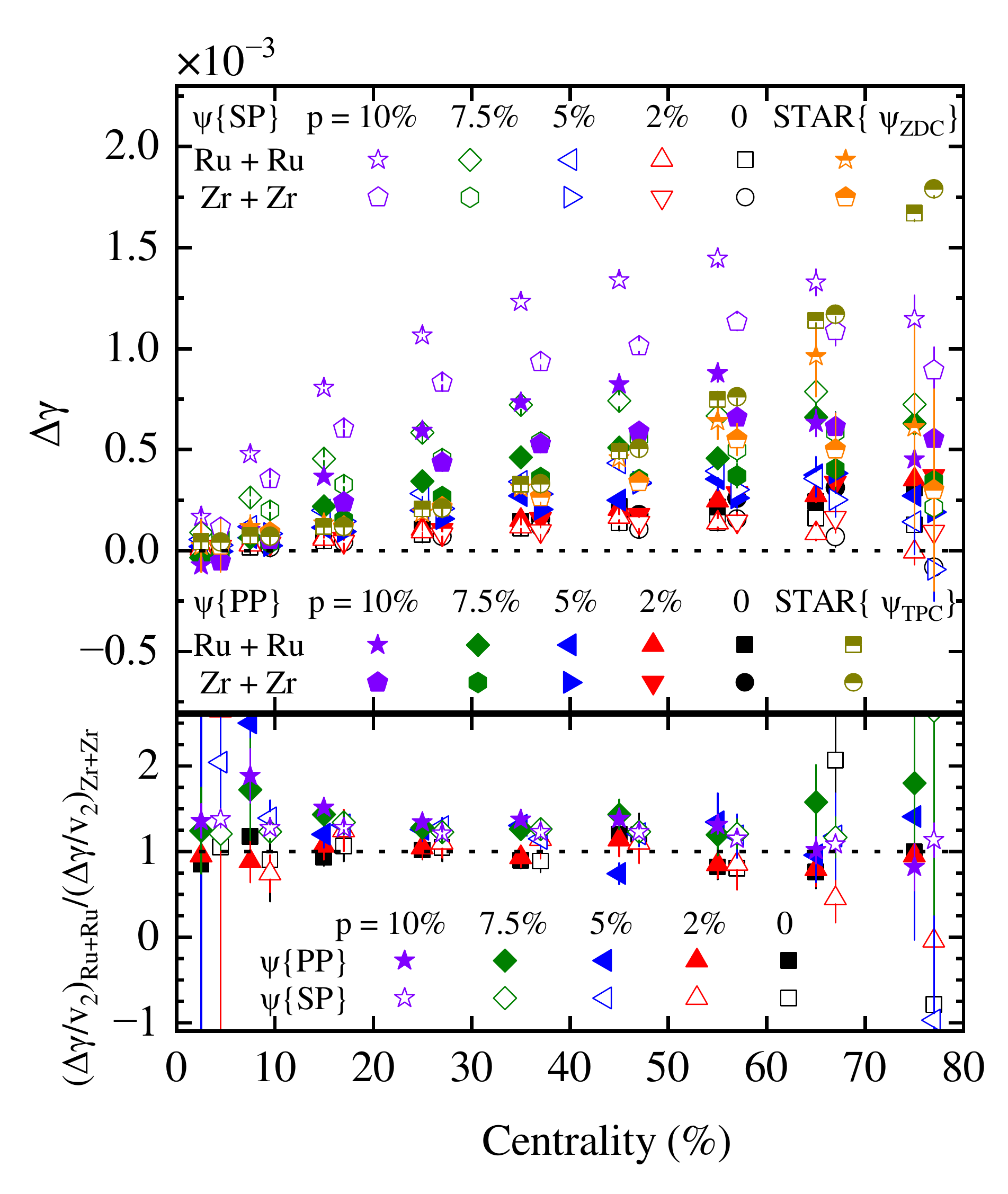}
	\caption{(Color online) Upper panel: The centrality dependences of $\Delta \gamma\{\mathrm{PP}\}$ (solid symbols) and $\Delta \gamma\{\mathrm{SP}\}$ (open symbols) in isobar collisions at $\sqrt{s_{_{\rm NN}}} = 200$ GeV from the AMPT model with different strengths of the CME, in comparison with the STAR data~\cite{STAR:2021mii}. Lower panel: The centrality dependences of $\Delta \gamma / \mathrm{v}_2$  ratios of Ru + Ru collisions to Zr + Zr collisions. The data points are shifted along the $x$ axis for clarity.
	}
	\label{fig-delgamma}
\end{figure}

The upper panel of Fig.~\ref{fig-delgamma} shows the centrality dependences of $\Delta \gamma\{\mathrm{PP}\}$ and $\Delta \gamma\{\mathrm{SP}\}$ in Ru + Ru and Zr + Zr collisions at $\sqrt{s_{_{\rm NN}}} = 200$ GeV from the AMPT model with different strengths of the CME. Compared to the STAR data, our results favor a small percentage of CME signal, which is also consistent with our recent study~\cite{Zhao:2022grq}. The $\Delta \gamma\{\mathrm{SP}\}$ is greater than $\Delta \gamma\{\mathrm{PP}\}$, which indicates that $\Delta \gamma\{\mathrm{SP}\}$ is more sensitive to the CME than $\Delta \gamma\{\mathrm{PP}\}$ because the spectator plane is more strongly correlated with the direction of the magnetic field than the participant plane. The lower panel in Fig.~\ref{fig-delgamma} shows the $\Delta \gamma / v_2$ ratios of Ru + Ru to Zr + Zr, which are consistent with unity within our statistical errors.

\begin{figure}
	\includegraphics[scale=0.40]{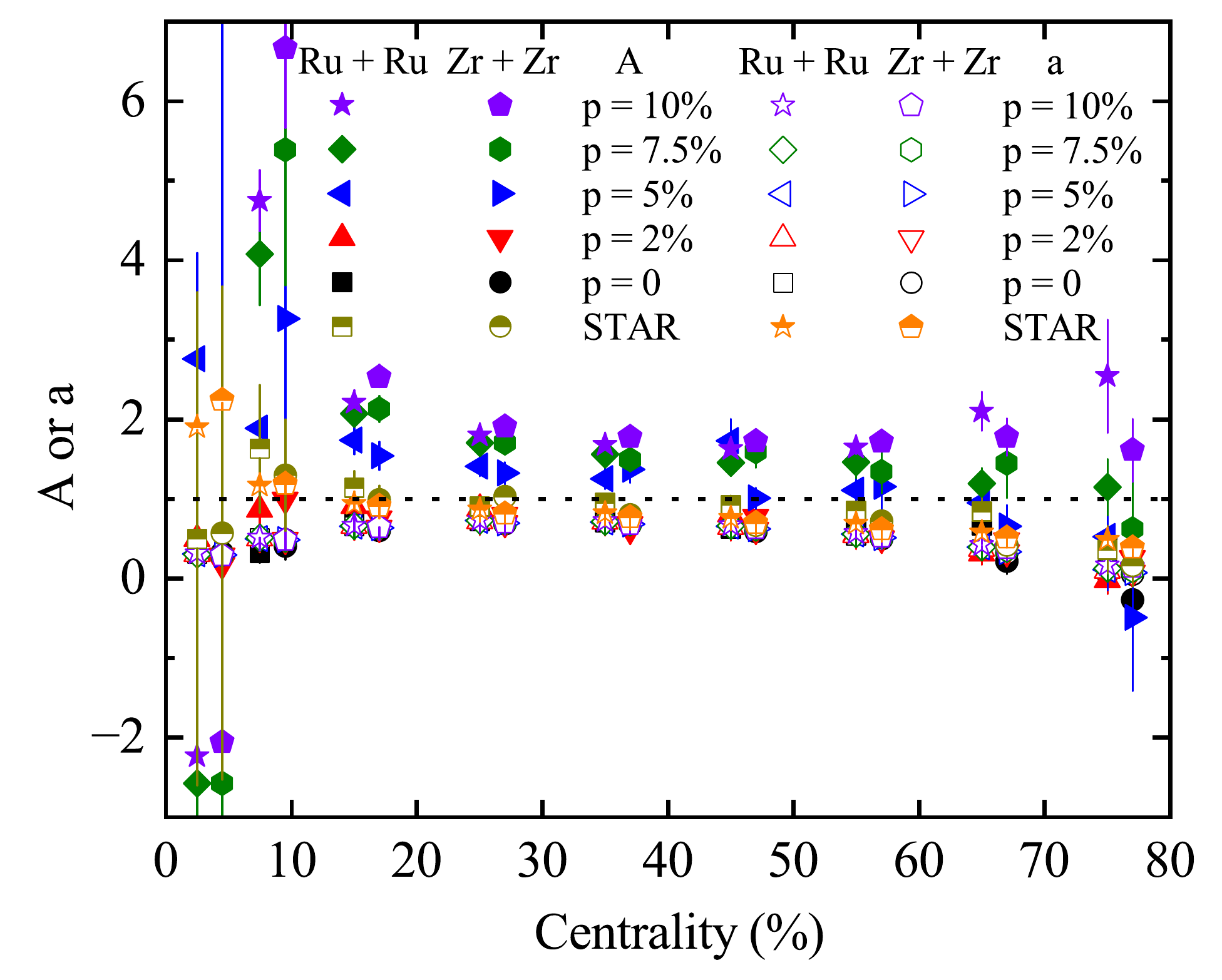}
	\caption{(Color online) The centrality dependences of $A$ and $a$ in isobar collisions at $\sqrt{s_{_{\rm NN}}} = 200$ GeV from the AMPT model with different strengths of the CME, in comparison with the STAR data~\cite{STAR:2021mii}. The solid and open symbols represent the results for $A$ and $a$, respectively. The data points are shifted along the $x$ axis for clarity.
	}
	\label{fig-Aora}
\end{figure}

Figure~\ref{fig-Aora} shows the centrality dependences of $A$ and $a$ from the AMPT model with different strengths of the CME, compared with the STAR experimental data~\cite{STAR:2021mii}.  As the CME strength increases, the value of $a$ hardly changes and is always less than unity. We have checked that $a$ satisfies the expectation relation of $a = \left\langle\cos 2\left(\Psi_{\mathrm{PP}}-\Psi_{\mathrm{SP}}\right)\right\rangle$, which indicates that the CME has the same effect on the elliptic flows relative to different planes. On the other hand, when $\rm p$=0 and $\rm p$=2\%, the value of $A$ is almost identical and less than unity. However, in the other cases, the value of $A$ is greater than unity and increases with increasing $p$. This indicates that the CME has different effects on $\Delta\gamma$ relative to different planes.

\begin{figure}
	\includegraphics[scale=0.40]{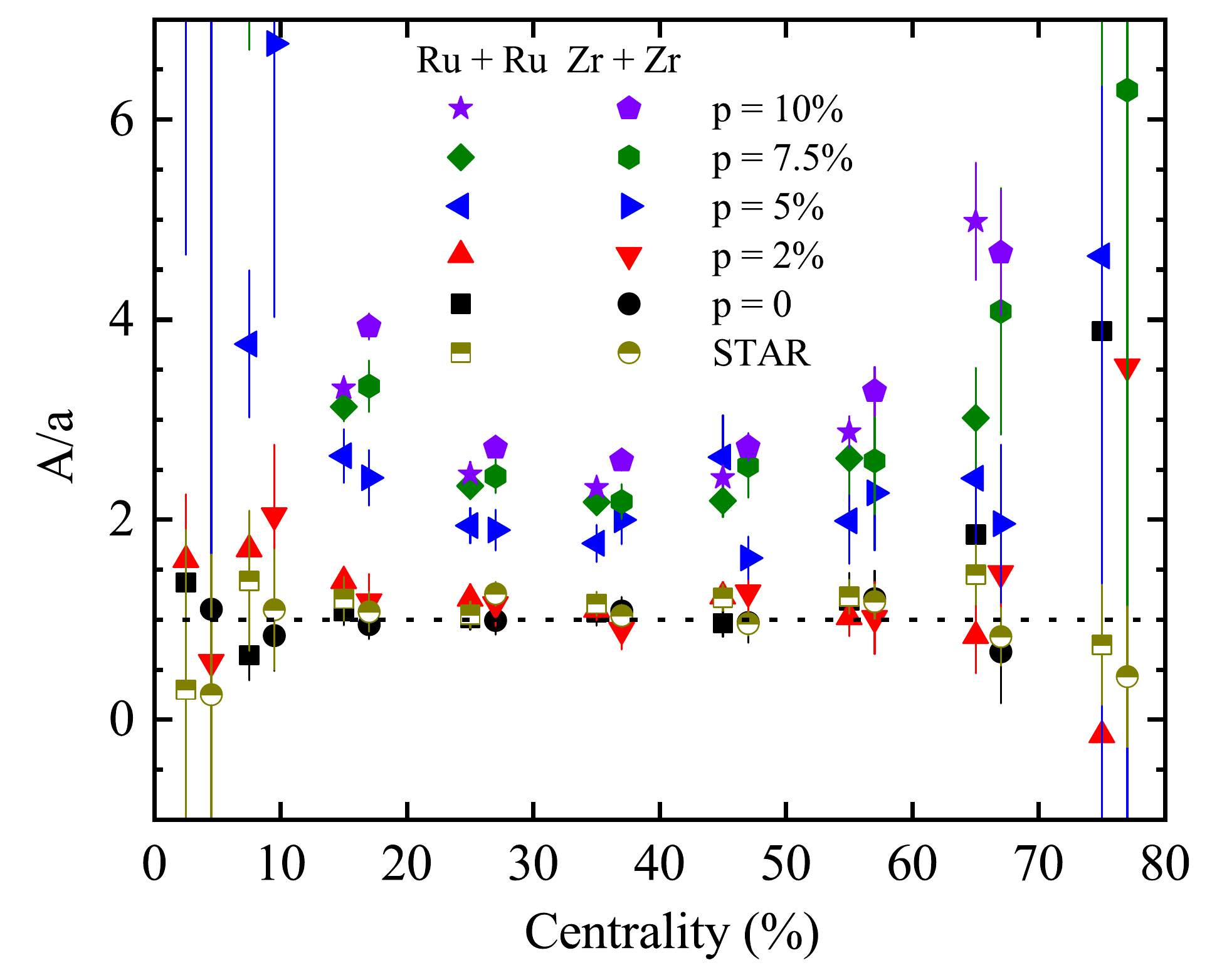}
	\caption{(Color online) The centrality dependences of A/a in isobar collisions at $\sqrt{s_{_{\rm NN}}} = 200$ GeV from the AMPT model with different strengths of the CME, in comparison with the STAR data~\cite{STAR:2021mii}. The data points are shifted along the $x$ axis for clarity.
	}
	\label{fig-Aratioa}
\end{figure}

Figure~\ref{fig-Aratioa} shows the $A/a$ ratio as a function of the centrality bin from the AMPT model with different strengths of the CME. A value of $A /a$ greater than one corresponds to the presence of a CME signal within the CME observable $\Delta\gamma$, according to  Eq.~(\ref{equ.09}) or Eq.~(\ref{equ.12}). Let us focus on mid-central centrality bins (20-50\%), where it is believed that the CME effect is more likely to be measured compared to other centrality bins. For the centrality bins of 20-50\%, we can clearly see that $A /a$ is greater than 1, except for the cases of $p$=0 and $p$=2\%. We also observe that the value of $A /a$ increases as the CME strength increases, suggesting that $A /a$ can reflect the strength of the CME signal.

\begin{figure}
	\includegraphics[scale=0.40]{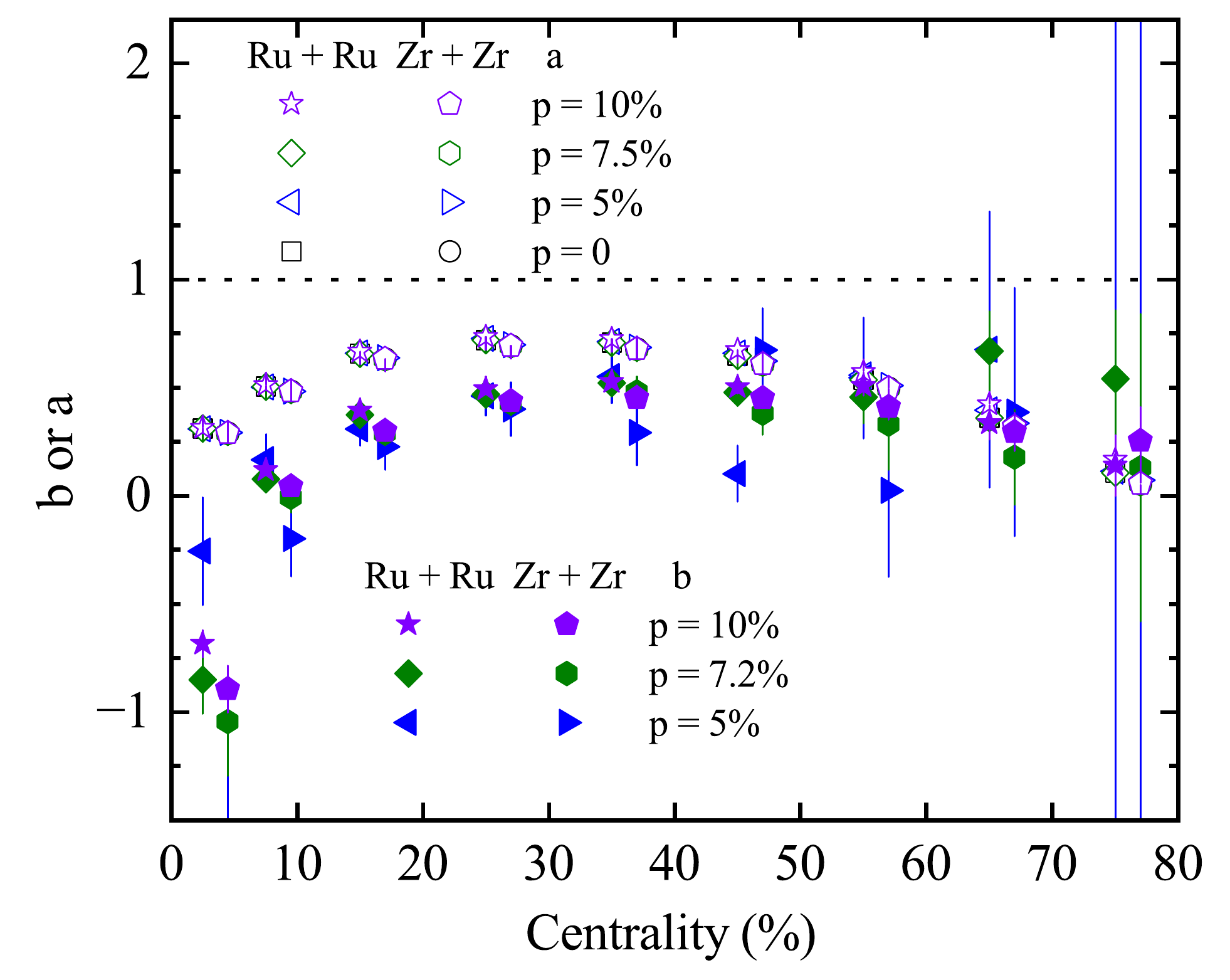}
	\caption{(Color online) The centrality dependences of $a$ and $b$ in isobar collisions at $\sqrt{s_{_{\rm NN}}} = 200$ GeV from the AMPT model with different strengths of the CME. The open and solid symbols represent the results for $a$ and $b$, respectively. The data points are shifted along the $x$ axis for clarity.
	}
	\label{fig-aorb}
\end{figure}

Next, we use Eq.~(\ref{equ.13})  to calculate $b$ and compare it with $a$. Figure \ref{fig-aorb} shows the centrality dependences of $a$ and $b$ from the AMPT model with different strengths of the CME. We observe that the values of $a$ and $b$ are different, where the value of $b$ is smaller than the value of $a$ for the centrality bins of 20-50\%. The $a$ value is almost independent of CME strength, consistent with that shown in Fig.~\ref{fig-Aora}. The value of $b$ indicates the relative ability of the PP method to carry the CME signal relative to the SP method. The value of $b$ does not vary much with the CME strength within the statistical errors. Note that we do not show the case where the CME strength is 2\% due to the huge statistical errors.

\begin{figure}
	\includegraphics[scale=0.40]{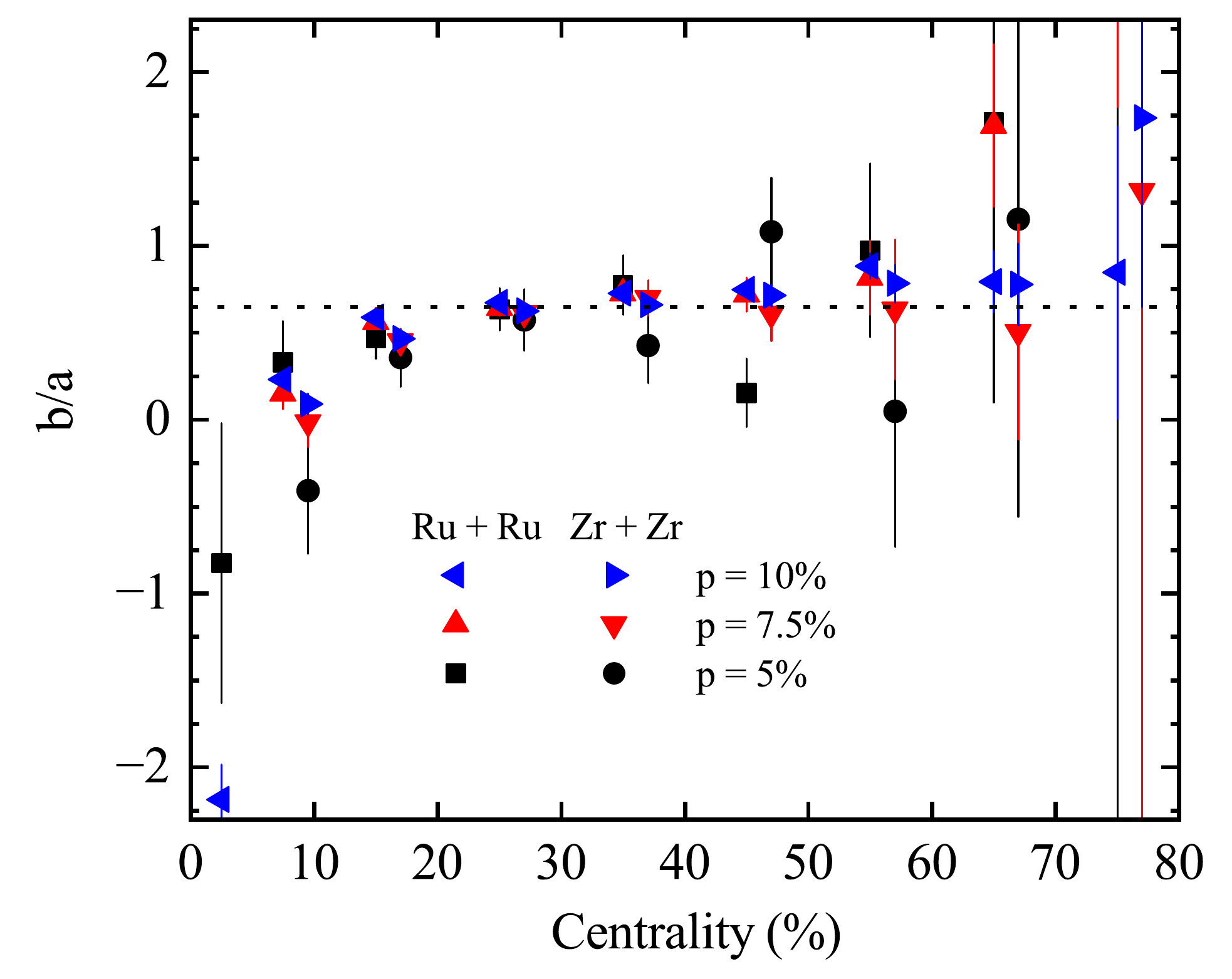}
	\caption{(Color online) The centrality dependences of $b/a$ in isobar collisions at $\sqrt{s_{_{\rm NN}}} = 200$ GeV from the AMPT model with different strengths of the CME. The data points are shifted along the $x$ axis for clarity.
	}
	\label{fig-bratioa}
\end{figure}

Figure~\ref{fig-bratioa} further shows the centrality dependences of the $b/a$ ratios from the AMPT model with different strengths of the CME. For the centrality bins of 20-50\%, an approximate relation of $b/a = 0.65( \pm 0.18)$ can be obtained by using a constant function fitting. It means that the relative ratio of the CME signals relative to different planes is not equal to the inverse of the ratio of the elliptic flows relative to different planes. Next, 
we will demonstrate the significance of this finding in determining the fraction of the CME signal within the $\Delta\gamma$ observable.
\begin{figure}
	\includegraphics[scale=0.4]{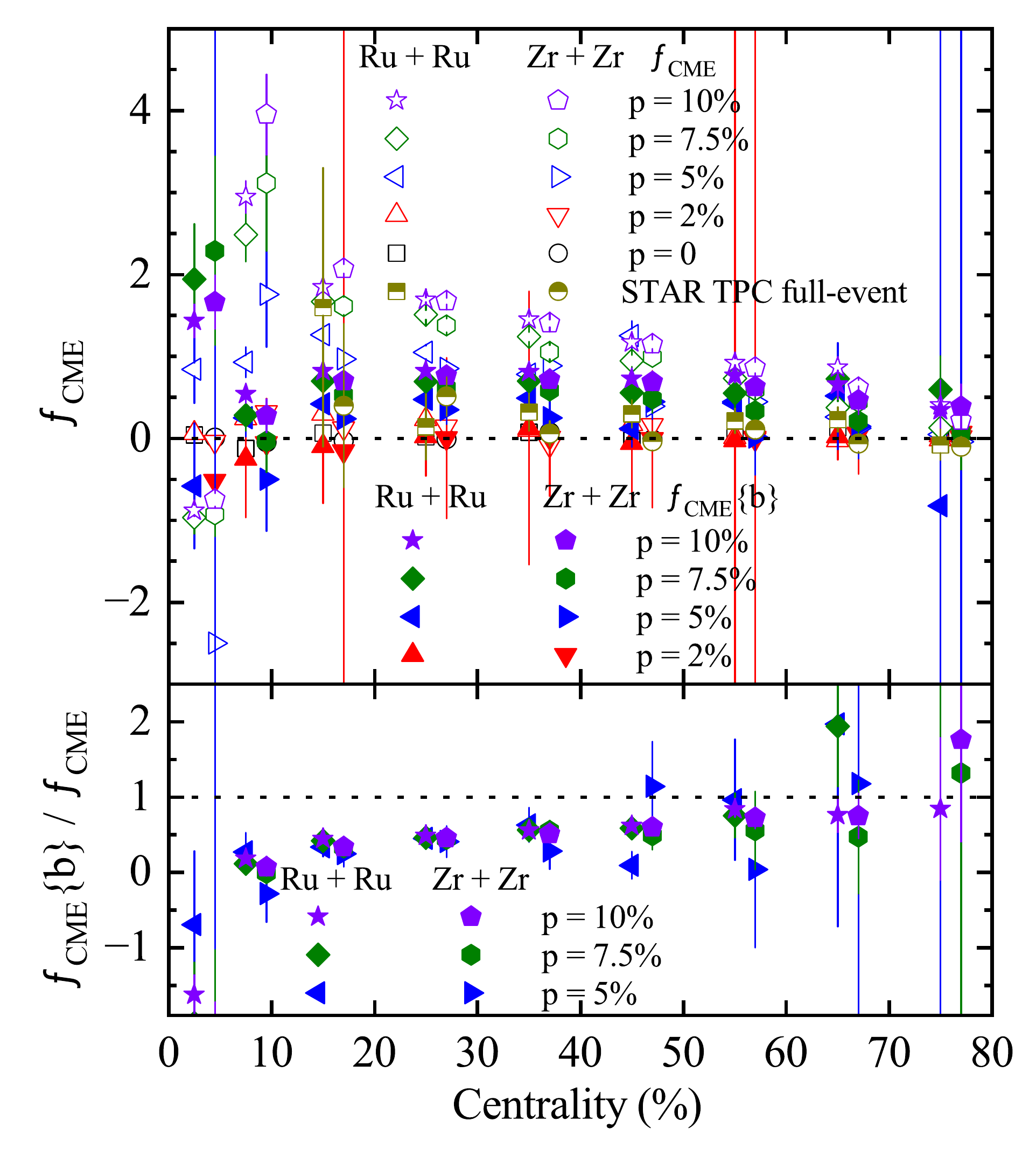}
	\caption{(Color online) Upper panel: The centrality dependences of $f_{\mathrm{CME}}\{b\}$ and $f_{\mathrm{CME}}$ in isobar collisions at $\sqrt{s_{_{\rm NN}}} = 200$ GeV from the AMPT model with different strengths of the CME, in comparison with the STAR data~\cite{STAR:2021mii}. The solid and open symbols represent the results for $f_{\mathrm{CME}}\{b\}$ and $f_{\mathrm{CME}}$, respectively. Lower panel: The centrality dependences of the ratio of $f_{\mathrm{CME}}\{b\}$ to $f_{\mathrm{CME}}$ corresponding to the upper panel. The data points are shifted along the $x$ axis for clarity.
	}
	\label{fig-fcme}
\end{figure}

The upper panel in Fig.~\ref{fig-fcme} shows two types of $f_{\mathrm{CME}}$ as a function of centrality from the AMPT model with different strengths of the CME, where the solid and open symbols represent $f_{\mathrm{CME}}$ calculated according to Eq.~(\ref{equ.09}) and  $f_{\mathrm{CME}}\{\mathrm{b}\}$ calculated according to Eq.~(\ref{equ.12}), respectively. Compared to the STAR experimental data~\cite{STAR:2021mii}, the results of $f_{\mathrm{CME}}$ from $p$=0\% or $p$=2\% are favored. For the centrality bins of 20-50\%,  $f_{\mathrm{CME}}\{b\}$ is less than $f_{\mathrm{CME}}$ when $p$ is not equal to zero. The lower panel in Fig.~\ref{fig-fcme}  shows the centrality dependence of the ratio of $f_{\mathrm{CME}}\{\mathrm{b}\} $ to $ f_{\mathrm{CME}}$. For the centrality bins of 20-50\%, the ratio is less than unity, which suggests that the fraction of the CME signal within the $\Delta\gamma$ observable will be overestimated if the relation of $b$ = $a$ is assumed.

\begin{figure}
	\includegraphics[scale=0.4]{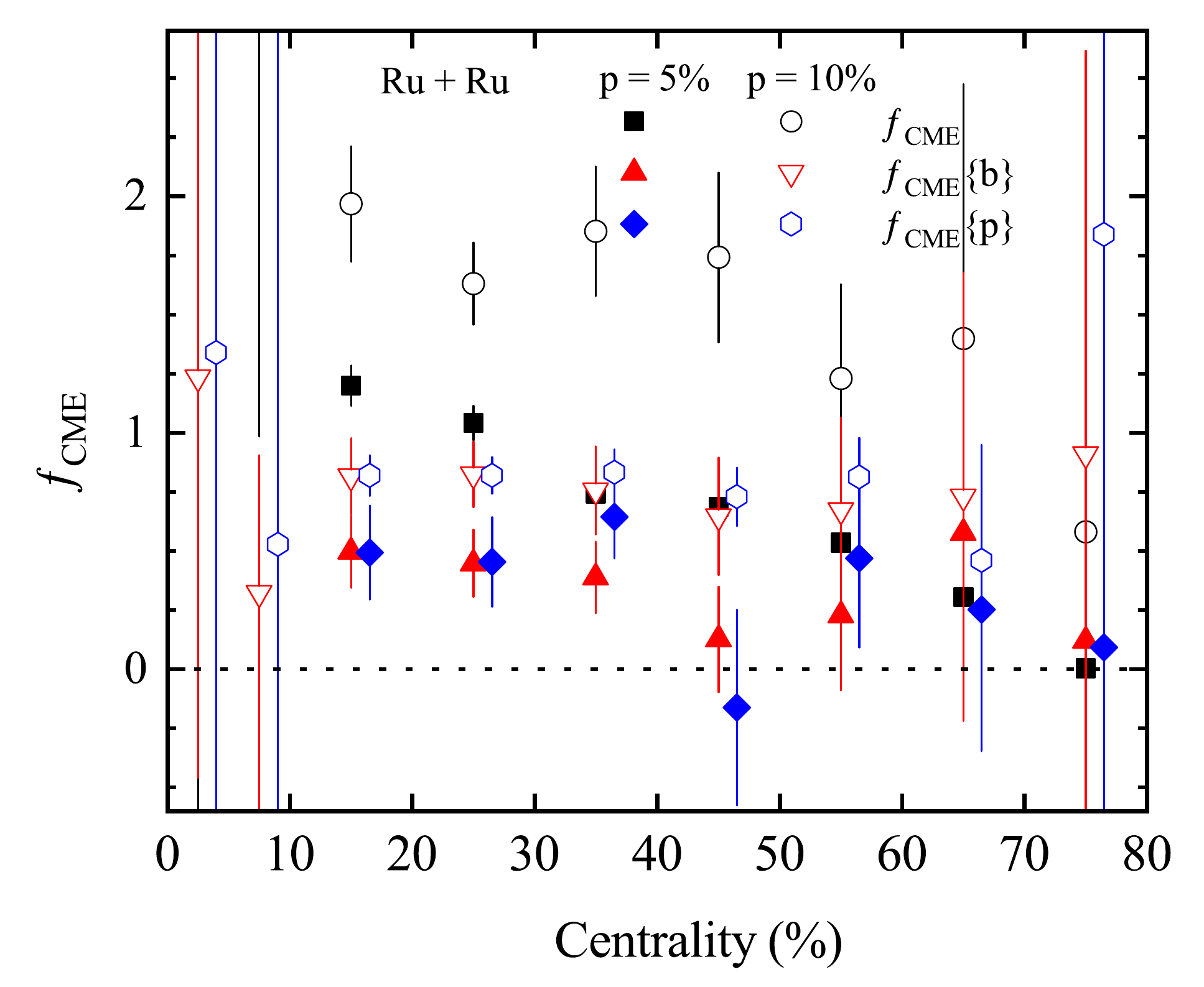}
	\caption{(Color online) The centrality dependences of  $f_{\mathrm{CME}}$, $f_{\mathrm{CME}}\{b\}$ and $f_{\mathrm{CME}}\{p\}$ from three different methods in Ru + Ru collisions at $\sqrt{s_{_{\rm NN}}} = 200$ GeV from the AMPT model with two different strengths of the CME. The data points are shifted along the $x$ axis for clarity. 
	}
	\label{fig-fcme3}
\end{figure}

To discern which of the two above $f_{\mathrm{CME}}$ is closer to the real situation, we can theoretically use another method to obtain the true $f_{\mathrm{CME}}$, namely $f_{\rm CME}\{p\}$, as a criterion. It is defined as 
\begin{equation}\label{equ.16}
	f_{\rm CME}\{p\}=\frac{\Delta \gamma_{\rm CME}\{\mathrm{PP}\}(p \neq 0)}{\Delta \gamma\{\mathrm{PP}\}(p \neq 0)},
\end{equation}
where 
\begin{equation}\label{equ.17}
\begin{split}
\Delta \gamma_{\rm CME}\{\mathrm{PP}\}(p \neq 0)
=\Delta \gamma\{\mathrm{PP}\}(p \neq 0) \\ -\Delta \gamma\{\mathrm{PP}\}(p=0).
\end{split}
\end{equation}
The observables of $\Delta \gamma\{\mathrm{PP}\}(p = 0)$ and $\Delta \gamma\{\mathrm{PP}\}(\mathrm{p}\neq0)$ can be obtained from the AMPT model without the CME and with different strengths of the CME, respectively. Figure~\ref{fig-fcme3} shows that $f_{\mathrm{CME}}\{\mathrm{b}\}$ is closer to $f_{\mathrm{CME}}\{\mathrm{p}\}$ than $f_{\mathrm{CME}}$ for the centrality bins of 10$-$50$\%$. This suggests that it is necessary to consider $b$ to obtain a more reliable $f_{\mathrm{CME}}$. We also calculated the $f_{\mathrm{CME}}\{\mathrm{p}\}$ relative to the SP, which is slightly larger than that relative to the PP.

\begin{figure}
	\includegraphics[scale=0.4]{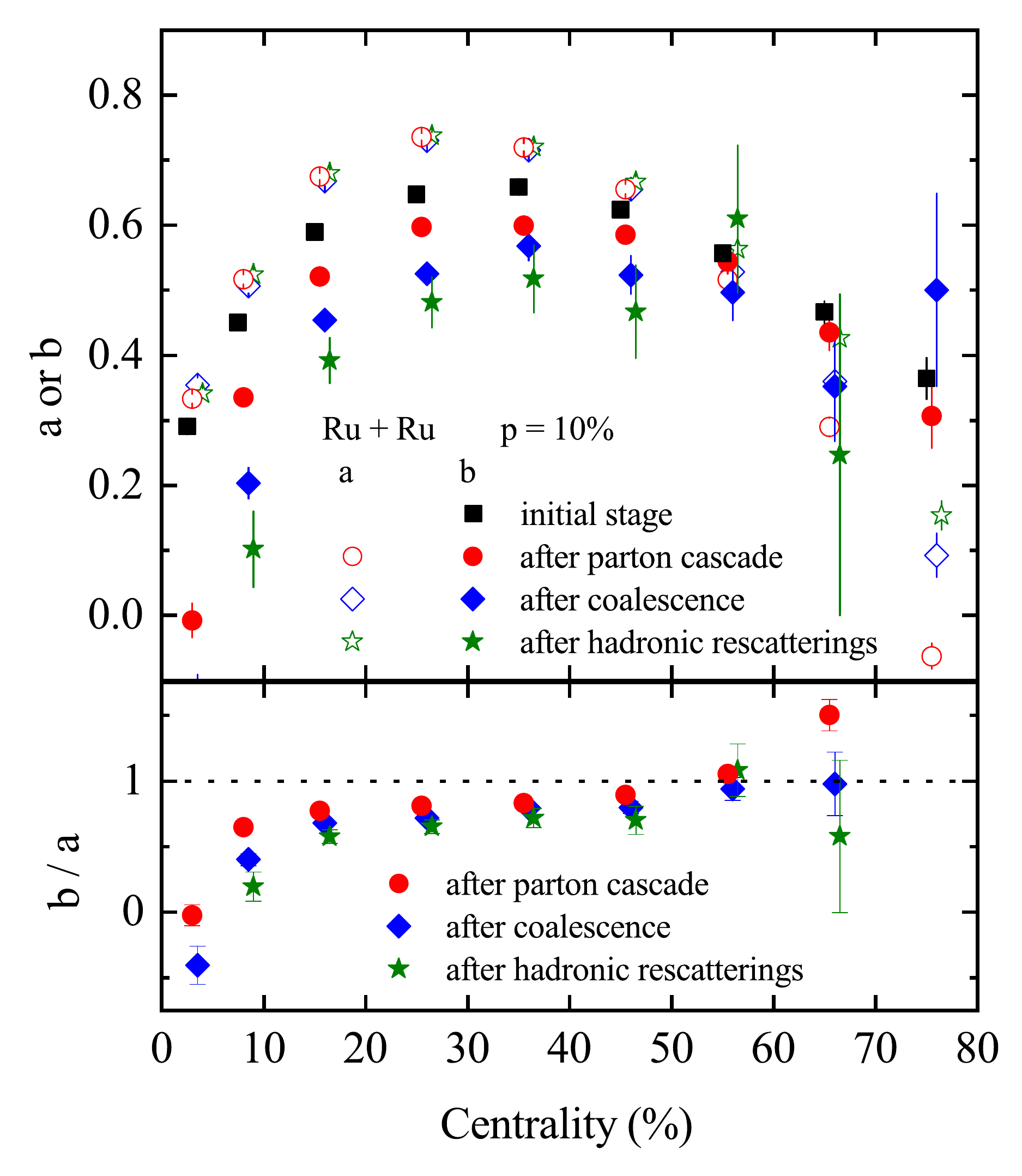}
	\caption{(Color online) Upper panel: The centrality dependences of $a$ and $b$ in Ru + Ru collisions at $\sqrt{s_{_{\rm NN}}} = 200$ GeV for four different stages from the AMPT model with the CME strength of $p=10$\%. The open and solid symbols represent the results for $a$ and $b$, respectively. Lower panel: The centrality dependences of the ratio of $b/a$ for the different stages in Ru + Ru collisions corresponding to the upper panel. The data points are shifted along the $x$ axis for clarity.
	}
	\label{fig-9}
\end{figure}

It is evident from the above results that $b \neq a$ has a significant impact on the result of $f_{\mathrm{CME}}$ in isobar collisions. In order to understand the origin of this inequality, we calculated $a$ and $b$ for different stages of Ru + Ru collisions for the AMPT model with the CME strength of $p=10$\%. According to the framework of the AMPT model, we focus on four evolution stages of heavy-ion collisions, which are the initial stage, after parton cascade, after coalescence, and after hadron rescatterings. In the upper panel of Fig.~\ref{fig-9}, we observe that the value of $a$ remains unchanged during the last three stages. Note that we do not show $a$ for the initial stage, since the elliptic flow is initially zero. However, for a given evolutionary stage, the value of $b$ is always smaller than the value of $a$ and decreases stage by stage in the centrality bins of 10-50\%. The lower panel of Fig.~\ref{fig-9} shows that the ratio of $b/a$ also decreases with the stage evolution. The main reason is of course due to the change in $b$. The decrease of $b$ indicates a decreasing correlation between the CME signals relative to the two different planes, which can be interpreted as a result of decorrelation due to final-state interactions during the evolution of heavy-ion collisions~\cite{Ma:2011uma,Huang:2019vfy,Huang:2022fgq,Zhao:2022grq}.

\begin{figure}
	\includegraphics[scale=0.40]{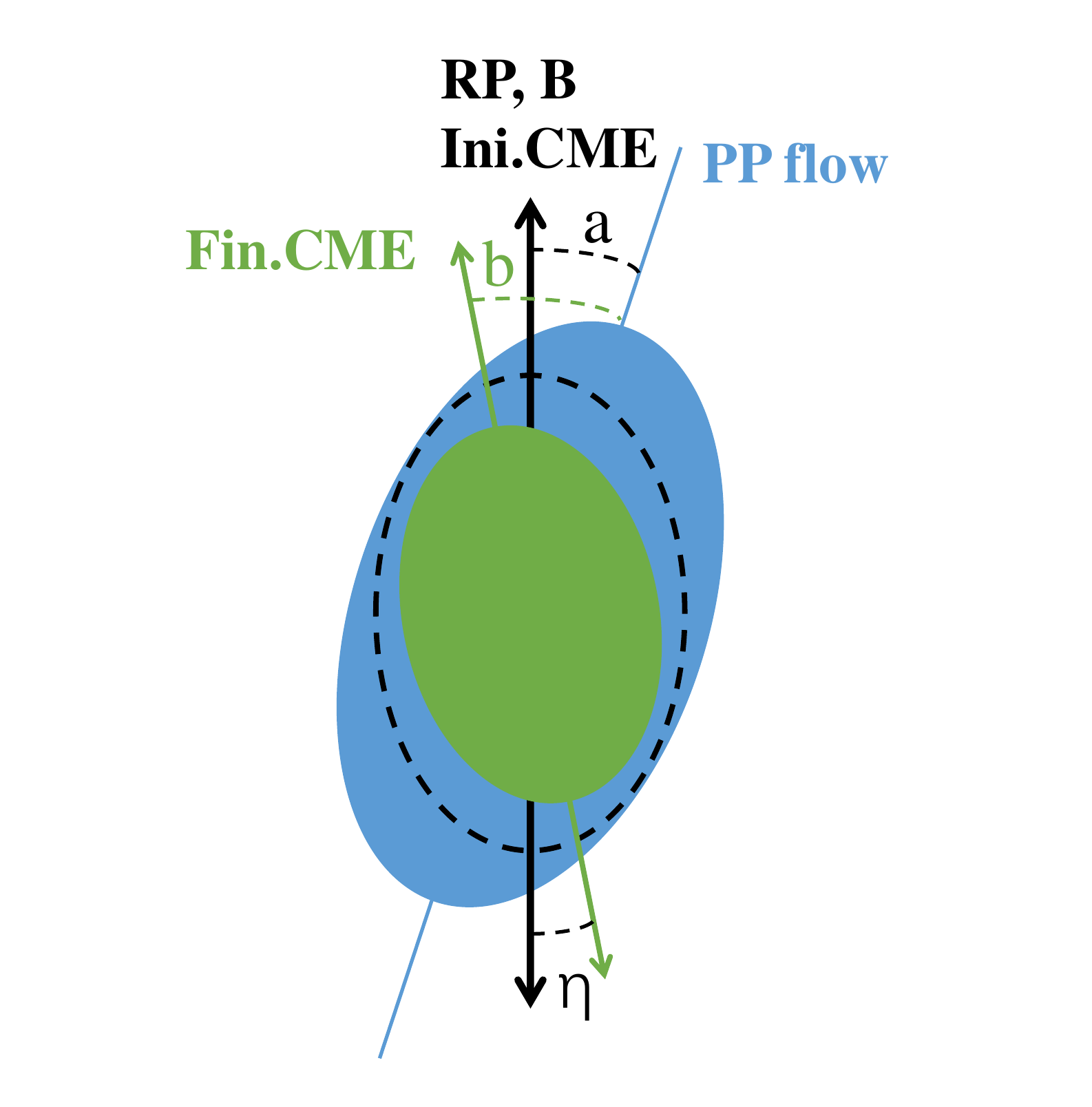}
	\caption{
	(Color online) An illustration of the change of the CME signal from the initial state to the final state in relativistic heavy-ion collisions. The initial CME signal is aligned with the magnetic field (black dashed ellipse) but at a different angle to the participant plane angle (blue ellipse). Final state interactions change the magnitude and direction of the CME signal, resulting in its reduction and rotation away from the initial direction, as shown by the green solid ellipse.
	}
	\label{fig10}
\end{figure}

A possible explanation for the difference between the $b$ and $a$ values is discussed below. In Ref.~\cite{Ma:2011uma}, the authors demonstrated that the final state interactions in relativistic heavy-ion collisions can significantly reduce the initial charge separation. The reduction factor can be as large as an order of magnitude. Because of the anisotropic overlap zone, the final state interactions not only reduce the magnitude but also alter the direction of the maximum CME current. As a result, the direction of the maximum survived signal deviates from the initial magnetic field direction (equivalently the spectator plane) by an angle (denoted as $\eta$). We illustrate this process with Fig.~\ref{fig10}~\cite{Fuqiang}. Since angular decorrelation also applies relative to the participant plane, one would naively expect $\frac{\Delta \gamma\{P P\}}{\Delta \gamma\{S P\}}=\frac{\langle\cos 2(\psi_{\rm PP}-\psi_{\rm SP}+\eta)\rangle}{\langle\cos 2 \eta\rangle}\approx\langle\cos 2(\psi_{\rm PP}-\psi_{\rm SP})\rangle$ to still hold. Our results indicate that this naive expectation does not seem to hold in AMPT, whose mechanism warrants further investigation. If $b$ is indeed smaller than $a$ as our AMPT study suggests, then the $f_{\mathrm{CME}}$ extracted from the SP/PP method assuming $b = a$ would be an overestimation of the final-state CME fraction in the measured $\Delta\gamma$ observable. However, since this decorrelation is a final-state reduction effect, the estimated $f_{\mathrm{CME}}$ would still be a lower limit of the initial CME signal that is subsequently damped and washed in the final-state evolution of relativistic heavy-ion collisions.

\section{Summary}
\label{sec:summary}
Using a multiphase transport model with different strengths of the CME, we reexamine the proposed two-plane method to determine the fraction of the CME signal within the CME observable of $\Delta\gamma$ in isobar collisions at $\sqrt{s_{_{\rm NN}}} = 200$ GeV. We first calculate the elliptic flow $v_2$ and the CME observables of $\Delta \gamma$ relative to the spectator plane and participant plane. The ratio $b$ of the CME signal relative to the two different planes is found to be different from the ratio $a$ of the background relative to the two different planes in isobar collisions. If the difference between $a$ and $b$ is taken into account, we demonstrate that it will lead to a smaller CME fraction than the constraint obtained in the current experimental way. We theoretically observe a decrease in the value of $b$ during the stage evolution in the AMPT model, which indicates the decorrelation of the chiral magnetic effect relative to spectator and participant planes is caused by final state interactions in isobar collisions. Since $a$ and $b$ were assumed to be equal in the current experimental study, the fraction of the final-state CME signal in $\Delta\gamma$ measurement could be overestimated. We are going to perform the calculations for Au + Au collisions at $\sqrt{s_{_{\rm NN}}} = 200$ GeV. We hope that our study will provide a theoretical reference for future measurements of the fraction of the chiral magnetic effect inside the experimental observable in relativistic heavy-ion collisions.

\section*{ACKNOWLEDGMENTS}
We thank Profs. Fuqiang Wang and Jie Zhao for their helpful discussions. This work is supported by the National Natural Science Foundation of China under Grants  No. 12325507, No.12147101, No. 11890714, No. 11835002, No. 11961131011, No. 11421505, No. 12105054, the National Key Research and Development Program of China under Grant No. 2022YFA1604900, the Strategic Priority Research Program of Chinese Academy of Sciences under Grant No. XDB34030000, and the Guangdong Major Project of Basic and Applied Basic Research under Grant No. 2020B0301030008.

\bibliography{fcme} 
 
\end{CJK*}
\end{document}